\documentclass[12pt]{article}
\usepackage[T1]{fontenc}
\usepackage{geometry}
\usepackage{amsmath}
\usepackage{graphics}
\usepackage{subfigure}

\makeatletter

\newcommand{\LyX}{L\kern-.1667em\lower.25em\hbox{Y}\kern-.125emX\spacefactor1000}
\let\SF@@footnote\footnote
\def\footnote{\ifx\protect\@typeset@protect
    \expandafter\SF@@footnote
  \else
    \expandafter\SF@gobble@opt
  \fi
}
\expandafter\def\csname SF@gobble@opt \endcsname{\@ifnextchar[
  \SF@gobble@twobracket
  \@gobble
}
\edef\SF@gobble@opt{\noexpand\protect
  \expandafter\noexpand\csname SF@gobble@opt \endcsname}
\def\SF@gobble@twobracket[#1]#2{}

\newcommand{\lyxaddress}[1]{
  \par {\raggedright #1 
  \vspace{1.4em}
  \noindent\par}
}

\usepackage[T1]{fontenc}


\begin{document}

\title{Electronic spectrum and hopping conductivity in highly doped lattice systems }

\author{Yuri G. Pogorelov, J. M. B. Lopes dos Santos, João M. V. P. Lopes}

\maketitle

\lyxaddress{Departamento de Física da Faculdade de Ciências da Universidade do Porto and
Centro de Física do Porto, 4150 Porto, Portugal}

\section{Introduction}

The interest to electronic processes in disordered systems was greatly inspired
by the fascinating disorder effects in semiconductors, including doped and amorphous
ones {[}1{]}, and in mesoscopic metallic systems {[}2{]}. Unlike the Bloch states
in fully periodic systems, the electronic spectrum of disordered systems generally
includes both extended and localized states, their coexistence being related
to the competition between kinetic and potential energy of Fermi particles.
The main consequence of this competition is the possibility for Anderson transition
from metallic to insulating state at zero temperature and sufficiently strong
disorder {[}3{]}. The best studied situation is that of non-interacting electrons
(that is, the single-electron approximation) in a certain random field. Efficiency
of single-electron theories for each type of electronic states in disordered
systems (at the Fermi level \( \epsilon _{F} \), they are mostly extended in
metals and mostly localized in semiconductors) is assured by the presence of
a certain small parameter, such as \( \lambda _{F}/\ell  \) (where \( \lambda _{F} \)
is the Fermi wavelength, \( \ell  \) the mean free path) in metals {[}2{]}
and \( na^{3} \) in semiconductors {[}4{]} (\( n \) is the concentration of
charge carriers, \( a \) the lattice parameter).

The following discussion is addressed to the doped semiconducting systems. In
traditional semiconductors, typical values of \( na^{3} \) do not exceed \( 10^{-4}\div 10^{-6} \).
This is determined by a very great localization radius \( r_{0}\sim (20\div 50)a \)
of a single localized shallow state and the Mott criterion for metallization
in the impurity band: \( nr_{0}^{3}\sim 0.02 \) {[}1{]}. At doping levels below
this value, the single-electron approach yeilds in a finite Fermi density of
(localized) states \( \nu _{F}=\nu (\epsilon _{F}) \) and in the Mott law for
hopping conductivity vs temperature \( T \): \( \sigma (T)\propto \{\exp \}(-BT^{-1/4})\textrm{ } \),
\( B\approx 2.1(r^{3}_{0}\nu _{F})^{-1/4} \) {[}5{]}. However, it was shown
by Efros and Shklovskii {[}6{]} that account of Coulomb interactions between
such shallow states leads to formation of a ``soft gap'' in the unperturbed
density of states \( \nu _{F} \) near the Fermi level so that: \( \nu (\epsilon -\epsilon _{F})^{2}/e^{6} \)
until \( |\epsilon -\epsilon _{F}|\sim \Delta =e^{3}\nu ^{1/2}_{F}/\kappa ^{3/2} \)
(where \( e \) is the electron charge and \( \kappa  \) the static dielectric
constant). Consequently, the Mott law is changed to: \( \sigma (T)\propto \exp (B'T^{-1/2}) \),
\( B'\approx e/(\kappa r_{0})^{1/2} \) at sufficiently low temperatures, \( T<T_{c}\approx e^{4}r_{0}\nu _{F}/\kappa ^{2} \).
The above theoretic dependencies are in a good agreement with the bulk of experimental
data in traditional semiconductors.

A special class of doped materials, displaying semiconducting, metallic, superconducting
and various magnetic phases, is comprised by the doped perovskite systems {[}7,8{]}.
From the point of view of standard theory of semiconductors, these materials
exhibit extremely high values of \( na^{3}\sim 0.1\div 0.5 \), that is more
than three orders of magnitude higher than those observed in traditional semiconductors.
So the experimentally observed hopping type of conductance, at sufficiently
high temperatures, in perovskite manganites Ln\( _{1-x} \)A\( _{x} \)MnO\( _{3} \)
(where Ln is a lanthanide, A an alkali-earth metal) with doping levels \( x\sim 0.3 \)
{[}9{]} is quite a striking fact. It indicates that Fermi states in such materials
remain localized even at so heavy doping and, from the before cited Mott criterion,
the upper limit for localization radius should be estimated as \( r_{0}\sim 0.4a \).
This is an opposite limit to the traditional case of shallow dopants, hence
a different evolution of the excitation spectrum can be expected. In particular,
the effects of electron-electron interaction can be much more pronounced.

Of course, real perovskite manganites possess many other peculiar properties,
as spin-dependent kinetic energy (it is just this dependence that suppresses
kinetic energy in the paramagnetic phase) and the related Zener mechanism of
double exchange {[}10{]}, strong coupling of charge carriers to Jahn-Teller
deformations {[}11{]}, possible formation of small spin polarons {[}12{]}, charge
localization {[}13{]} and charge ordering {[}14{]}, etc. However all the above
mechanisms are usually discussed within uniform (that is, completely ordered)
models, while our main focus now is on the effects specific for extremely strong
disorder. To this end, we propose a model approximation, starting from a set
of strictly localized single-site electronic states, with random energy in each
site formed by the superposition of classic Coulomb potentials from other (occupied)
sites and from the fixed charged dopants (so that the overall electroneutrality
is assured). At the next step, the kinetic energy is considered as a small perturbation,
triggering the hops between the nearest neighbor occupied and empty sites (if
the necessary energy difference is compensated by phonons, a.c. electric fields,
etc.). This model is interesting firstly as an opposite limit to the usual situation
when the disorder is treated as a small perturbation of an initially uniform
system, and secondly as a new realization of strongly correlated many-body system.

Below in Sec. 2 we define the model parameters and describe the numeric processes
to seek the ground state and analyze various types of excitation spectra at
zero temperature, as functions of system size \( L \) and doping concentration
\( x \). Then the size, shape and topology independent behavior of the spectra
is rapidly attained with growing \( L \) (already at \( L=10\div 12 \) cell
units), and the main findings are:

\begin{description}
\item [(a)]a non-ergodicity of the full phase space with a definite distribution of
local energy minima and practically identical excitation spectra with respect
to all typical local minima; 
\item [(b)]asymmetric deviations from the mean-field \( \propto (\epsilon -\epsilon _{F})^{2} \)
behavior for the density of single-particle excitations; 
\item [(c)]vanishing of the density of pair excitations in the limit of low excitation
energy; 
\item [(d)]finiteness of the total density of many-body excitations in the same limit. 
\end{description}
The further development of the model in Sec. 3 involves finite temperatures
and hopping dynamics through the usual mechanism of deformation potential for
electron-phonon coupling. Then the non-ergodic structure of the system phase
space at zero temperature leads to its non-Markovian kinetics at finite temperatures.
A special numeric algorithm is developed, simulating this statistical process
and giving the temperature behavior of electronic specific heat, relaxation
times and diffusion coefficient (the latter being related to the conductivity
via the Einstein relation). In particular, for the typical concentration value
\( x=1/3 \) we found: \( \sigma (T)\propto \exp (-bT^{-\alpha }) \), with
\( \alpha \approx 0.87 \), different from the Efros and Shklovskii value 1/2.
The conclusions and perspectives for the further studies of the present model
are discussed in Sec. 4.

\section{Description of model. Zero Temperature}

The crystalline structure of lanthanum manganite LaMnO\( _{3} \) is close to
ideal perovskite with the (paramagnetic) lattice parameter \( a\approx 3.9 \)\AA
(Fig. 1a). Substitution of trivalent La\( ^{3+} \) by divalent alcali-earth
ion (say, Ca\( ^{2+} \)) brings an extra hole to the system which resides on
a nearby manganese site, changing its state from Mn\( ^{3+} \) to Mn\( ^{4+} \).
For a single dopant in the lattice, an arbitrarily weak tunneling will be sufficient
to produce the hole state equally shared between eight Mn sites, nearest neighbors
to the dopant Ca. However, for a finite doping, there appear random energy differences
between these sites, and if these differences are greater than the tunneling
amplitude (the kinetic energy), the hole will mainly occupy the lowest energy
site. The localization is also favoured here by the suppression of kinetic energy:
firstly due to the presence of intercalating oxygens between manganese sites
and secondly due to the incoherence of manganese spins at higher temperatures.

Referring to this situation we consider the model where the dopant ions occupy
randomly the central sites in the simple cubic lattice with probability \( x<1 \)
(Fig.~1b). Each dopant releases one charge carrier into the crystal and thus
acquires a unit\begin{figure}
{\centering \resizebox*{15cm}{!}{\includegraphics{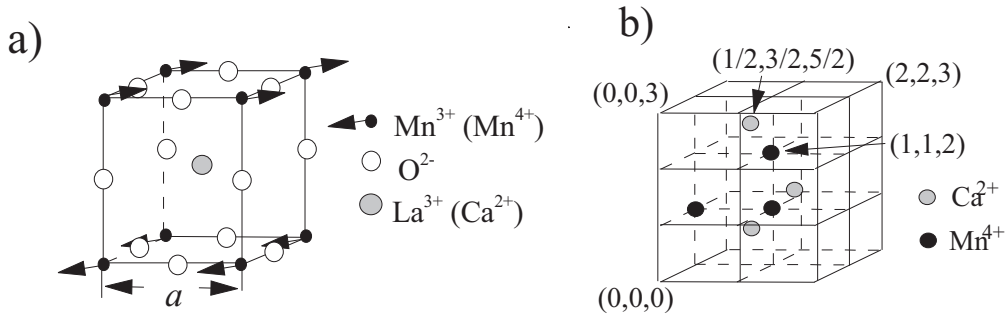}} \par}

\caption{\label{fig1} a) Elementary cubic cell corresponding to the LaMnO\protect\( _{3}\protect \)
structure. b) A \protect\( 2\times 2\times 3\protect \) parallelepiped sample
of cubic lattice with a random distribution of charged dopants and charge carriers
at concentration \protect\( x=1/4\protect \).}
\end{figure} charge \( e \) of opposite sign. In neglectance of hops between lattice sites,
each carrier occupy single lattice site and the total electronic energy includes
only Coulomb contributions:

\begin{equation}
\label{eq1}
E_{c}=\sum _{{\bf n}}c({\bf n})\left[ \frac{1}{2}U({\bf n})-V({\bf n})\right] ,
\end{equation}
 where \( U({\bf n})=\sum _{{\bf n'}\neq {\bf n}}c({\bf n})u(|{\bf n}-{\bf n'}|) \),
\( V({\bf n'})=\sum _{{\bf n'}\neq {\bf n}}c({\bf n'})u(|{\bf n}-{\bf n'}-\delta |) \),
\( u(r)=e^{2}/\kappa r \), \( c({\bf n}) \) is an occupation number for charge
carrier in the lattice site \( {\bf n}=a(n_{1},n_{2},n_{3}) \) (with integer
\( n_{i} \)), while occupation of the dopant site \( {\bf n+\delta } \), with
\( {\bf \delta }=a(\frac{1}{2},\frac{1}{2},\frac{1}{2}) \), in the same cell
is defined by \( d({\bf n}) \). In what follows we consider fixed random configuration
\( d({\bf n}) \) of ``frozen'' dopants, then the system ground state corresponds
to such adjustment of the configuration \( c({\bf n}) \) of carriers that \( E_{c} \)
is a minimum.

For arbitrary configuration \( c({\bf n}) \), when a carrier is taken off from
the site \( \bf n \) (if occupied) or put into this site (if empty), the full
energy respectively decreases or increases by \( \epsilon ({\bf n})=U({\bf n})-V({\bf n}) \),
which can be thus associated with single-particle excitations of the Fermi liquid
theory. Here both \( U({\bf n}) \) and \( V({\bf n}) \) are random, but only
\( V({\bf n}) \), determined by the fixed configuration of dopants, can be
considered an usual local random field of Anderson's model {[}3{]} whereas \( U({\bf n}) \),
determined by the variable configuration of carriers, is substantially non-local.
Hence each \( \epsilon ({\bf n}) \) essentially depends on the positions of
all other carriers, and the total \( E_{c} \) strongly differs from the sum
of all \( \epsilon ({\bf n}) \). In this situation, there is no evidence for
unique energy minimum and the structure of phase space can be very complicate.
However, the consideration is simplified if one takes in mind that any two configurations
in this space, satisfying the same normalization condition, Eq. (2), can be
connected by a sequence (a phase trajectory) of single-particle moves between
nearest neighbor occupied and empty sites. In fact, the following treatment
is limited just to this class of trajectories.

Besides the above indicated single-particle energies \( \epsilon (\bf n) \),
the excitation spectrum includes also the so-called pair energies {[}6{]}:
\[
\epsilon ({\bf n},{\bf n'})=\epsilon ({\bf n'})-\epsilon ({\bf n})-u'(|{\bf n}-{\bf n'}|),\]
 that is the energy change at moving a carrier from the occupied site \( \bf n \)
to empty site \( \bf n' \). The last, ``excitonic'' term in this expression,
accounting for the correlation between different single-site energies, is just
responsible for opening of the Coulomb gap in the single-particle spectrum \( \nu _{(s-p)}(\epsilon ({\bf n})) \).
If the correlations between different pair energies \( \epsilon ({\bf n},{\bf n'}) \)
and \( \epsilon ({\bf n'},{\bf n''}) \) are neglected, which corresponds to
the mean-field approximation, a conclusion can be drawn that the pair spectrum
\( \nu _{p}(\epsilon ({\bf n},{\bf n'})) \) is ungapped {[}6{]}. But, as will
be seen from the exact numeric analysis below, in fact the density of such excitations
also vanishes at \( \epsilon \rightarrow 0 \), and this situation may be supposed
to exist for higher order excitations as well.

In our numeric procedure we consider lattice samples in the form of finite parallelepipeds
and apply the following algorithm. The initial configurations \( d_{0}(\{{\bf n}\}) \)
and \( c_{0}(\{{\bf n}\}) \) are defined by assigning them independently random
values 0 or 1 for each site \( \bf n \), so that the normalization condition
holds: 
\begin{equation}
\label{eq2}
\sum ^{L_{1}}_{n_{1}=1}\sum _{n_{2=1}}^{L_{2}}\sum _{n_{3=1}}^{L_{3}}d(n_{1}+\frac{1}{2},n_{2}+\frac{1}{2},n_{3}+\frac{1}{2})=\sum ^{L_{1}+1}_{n_{1}=1}\sum _{n_{2=1}}^{L_{2}+1}\sum _{n_{3=1}}^{L_{3}+1}c(n_{1},n_{2},n_{3})=N
\end{equation}
 related to the doping level \( x \) through \( N=\left[ xL_{1}L_{2}L_{3}\right]  \).
Then we choose, from all the \emph{nearest neighbor} pairs of occupied sites
\( \bf n \) and empty sites \( \bf n' \) , the pair \( \bf n_{0} \) and \( \bf n_{0}' \)
corresponding to the minimum value of pair energy: \( \epsilon ^{(neighb)}_{min}=\epsilon ({\bf n_{0}},{\bf n_{0}'}) \).
If \( \epsilon ^{(neighb)}_{min} \) is negative, we change from the configuration
\( c_{0}(\{{\bf n}\}) \) to a new configuration \( c_{1}(\{{\bf n}\}) \),
moving the carrier from \( \bf n_{0} \) to \( \bf n_{0}' \). This process
of single-particle moves is repeated \( m \) times, until we come to such a
configuration \( c_{m}(\{{\bf n}\}) \) that \( \epsilon ^{(neighb)}_{min} \)
is already positive. Then we search for the minimum \( \epsilon ^{(all)}_{min} \)
of \( \epsilon ({\bf n},{\bf n'}) \) over \emph{all} occupied \( \bf n \)
and empty \( \bf n' \) in this configuration, and, if it is negative, perform
the corresponding move. This procedure is repeated until such a configuration
\( c_{M}(\{{\bf n}\}) \) is reached that \( \epsilon ^{(all)}_{min} \) is
positive. Then \( c_{M}(\{{\bf n}\}) \) corresponds to a local equilibrium
(with respect to single-particle moves) and the respective value of \( E_{c}=E_{min}[c_{0}(\{{\bf n}\})] \),
a functional of the initial configuration, realizes a local minimum of energy.

Next, the system is ``shaken up'', that is, for the same initial dopant configuration
\( d(\{{\bf n}\}) \), a new arbitrary initial configuration \( c_{0}'(\{{\bf n}\}) \)
is created, restricted again by the normalization condition, Eq. (\ref{eq2}).
Then it is found that, with the same equilibration process, some new local equilibrium
\( c_{M'}(\{{\bf n}\}) \) and a new local minimum \( E_{min}'[c_{0}'(\{{\bf n}\})] \)
are obtained. The presence of various local minima is indicative of a non-ergodic
structure of the phase space (with a discrete topology restricted to single-particle
moves). Since there is a one-to-one correspondence between the initial state
and the final state of local equilibrium, the whole phase space of the system
gets divided into a number of attraction domains, each corresponding to a definite
local minimum. The absolute energy minimum, corresponding to the ``true''
ground state, can be defined as:
\begin{equation}
\label{eq3}
E^{abs}_{min}=\min _{c_{0}(\{{\bf n}\})}E_{min}[c_{0}(\{{\bf n}\})].
\end{equation}
 The consequent ``shake ups'' and equilibrations define a sort of Monte-Carlo
process to approach the ground state and this numeric process can be completed
within a reasonable time for not too big system (this is evidenced by the uniqueness
of the corresponding configuration \( c^{abs}_{M}(\{{\bf n}\}) \)). \begin{figure}
{\raggedright \subfigure[ ]{\resizebox*{!}{5cm}{\includegraphics{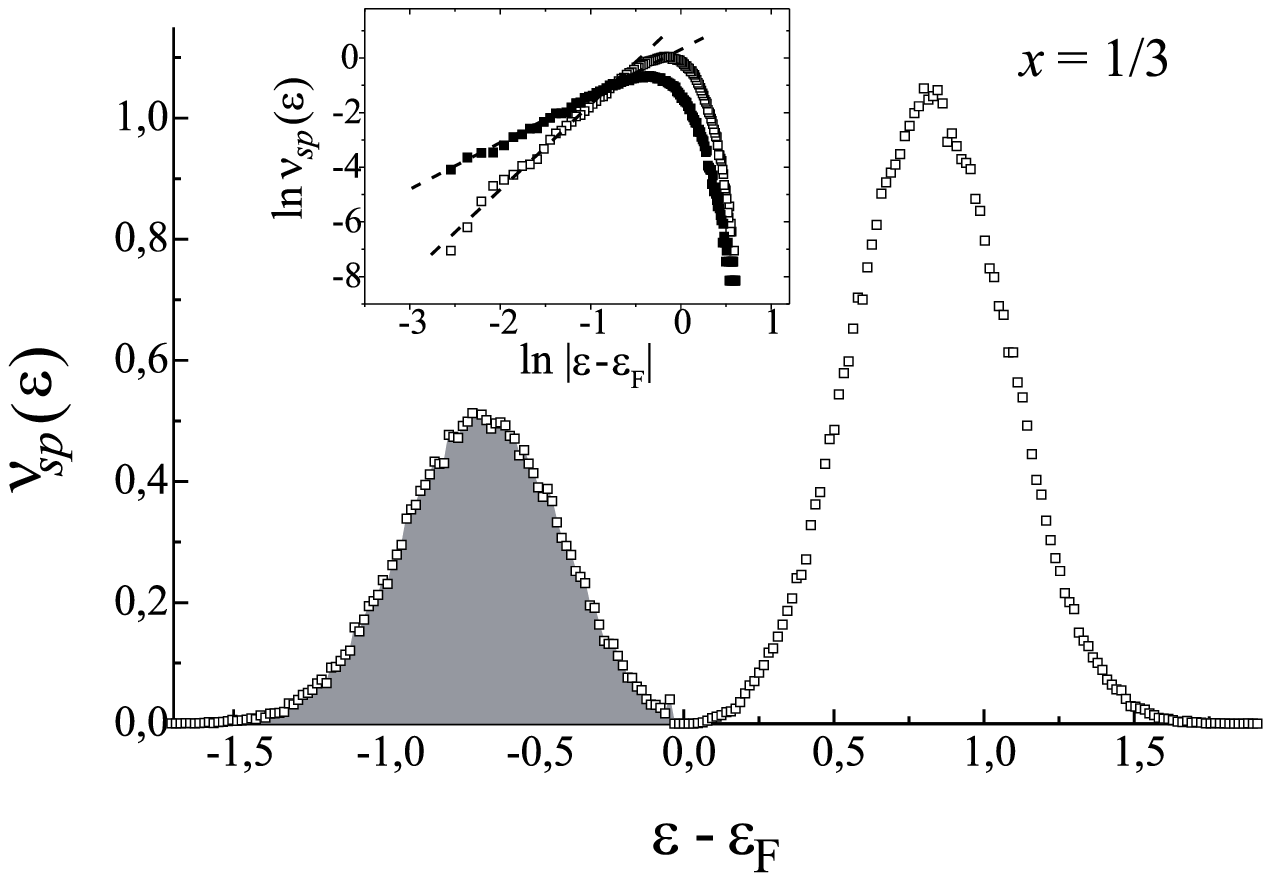}}} \par}

\vspace{-6.5cm}

{\raggedleft \subfigure[ ]{\resizebox*{!}{5cm}{\includegraphics{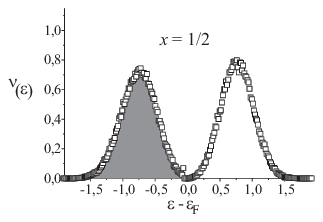}}} \par}

\caption{\label{fig2} a) Single-particle density of states at \protect\( T=0\protect \)
and doping level \protect\( x=1/3\protect \). Insert demonstrates different
power laws for densities of occupied (solid squares) and empty (light squares)
states close to Fermi level and their common Gaussian asymptotics far from it.
b) Single-particle density of states at \protect\( T=0\protect \) and \protect\( x=1/2\protect \).}
\end{figure} As a co-product, the process also provides the ``spectrum'' of local minima
\( \nu _{loc}(E)=\langle \delta (E-E_{min}[c_{0}])\rangle _{c_{0}} \), a new
characteristics of the strongly interacting disordered system. At a given dopant
configuration \( d(\{{\bf n}\}) \), the single-particle and the pair spectra,
\( \nu _{s-p}(\epsilon ) \) and \( \nu _{p}(\epsilon ) \), are calculated
with respect to each local minimum, including the true ground state. Within
accuracy to statistical noise, there is no difference found among the curves
taken with respect to different minima. This shows that the ``true'' ground
state is not any outstanding point between other equilibrium points. Finally
all the results are averaged over various dopant configurations \( d(\{{\bf n}\}) \).
For numeric simulations of the system, Eqs. (\ref{eq1}, \ref{eq2}), we began
from cubic samples of increasing size \( L \), then a size independent behavior
corresponding to thermodynamic limit is reached already at \( L\sim 10 \),
and these averaged characteristics are presented in Figs. 2-5.

The single-particle spectra \( \nu _{s-p}(\epsilon ) \) at different dopings,
shown in Fig. 2a,b, reveal a well-defined Coulomb gap around Fermi energy \( \epsilon _{F} \),
while their asymptotics far from \( \epsilon _{F} \) is well described by the
Gaussian law: \( \nu _{s-p}(\epsilon )\propto \exp \left[ -\alpha (\epsilon -\epsilon _{F})^{2}\right]  \),
in agreement with the known results of Lifshitz's theory of optimal fluctuation
for ``tail'' states in disordered systems {[}16{]}.

\begin{figure}
{\centering \resizebox*{10cm}{!}{\includegraphics{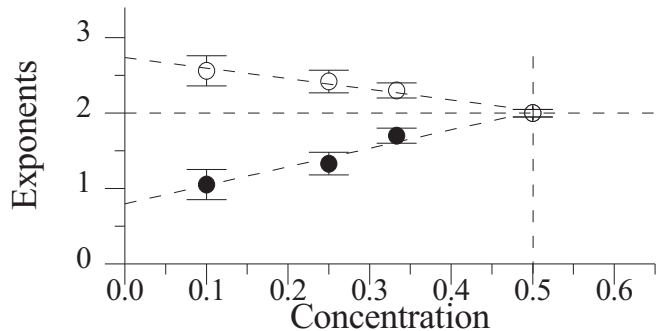}} \par}

\caption{\label{fig3} Power law exponents for densities of occupied (filled circles)
and empty (light circles) single-particle states as functions of concentration
of dopants.}
\end{figure}

A new notable feature of these spectra is a pronounced asymmetry between the
densities of empty and occupied states near \( \epsilon _{F} \), which are
described by different power laws:
\begin{equation}
\label{eq4}
\nu _{s-p}(\epsilon )\propto \left\{ \begin{array}{cc}
(\epsilon -\epsilon _{F})^{2+\eta '}, & \epsilon >\epsilon _{F,}\\
(\epsilon -\epsilon _{F})^{2-\eta ''}, & \epsilon <\epsilon .
\end{array}\right. 
\end{equation}
 Here the asymmetry factors \( \eta ' \) and \( \eta '' \) are not universal,
but vary with the doping and tend to zero at \( x\rightarrow 1/2 \) (Fig. \ref{fig3}).
The latter fact can be easily understood as a consequence of the symmetry between
filled and empty sites at this concentration.

Hence the mean-field quadratic law is found to be exact only at half-filling
(though the non-universal corrections to it, due to the higher order correlations,
can be really small in the case of traditional doped semiconductors with \( r_{0}\gg a \)).

\begin{figure}
{\centering \resizebox*{12cm}{!}{\includegraphics{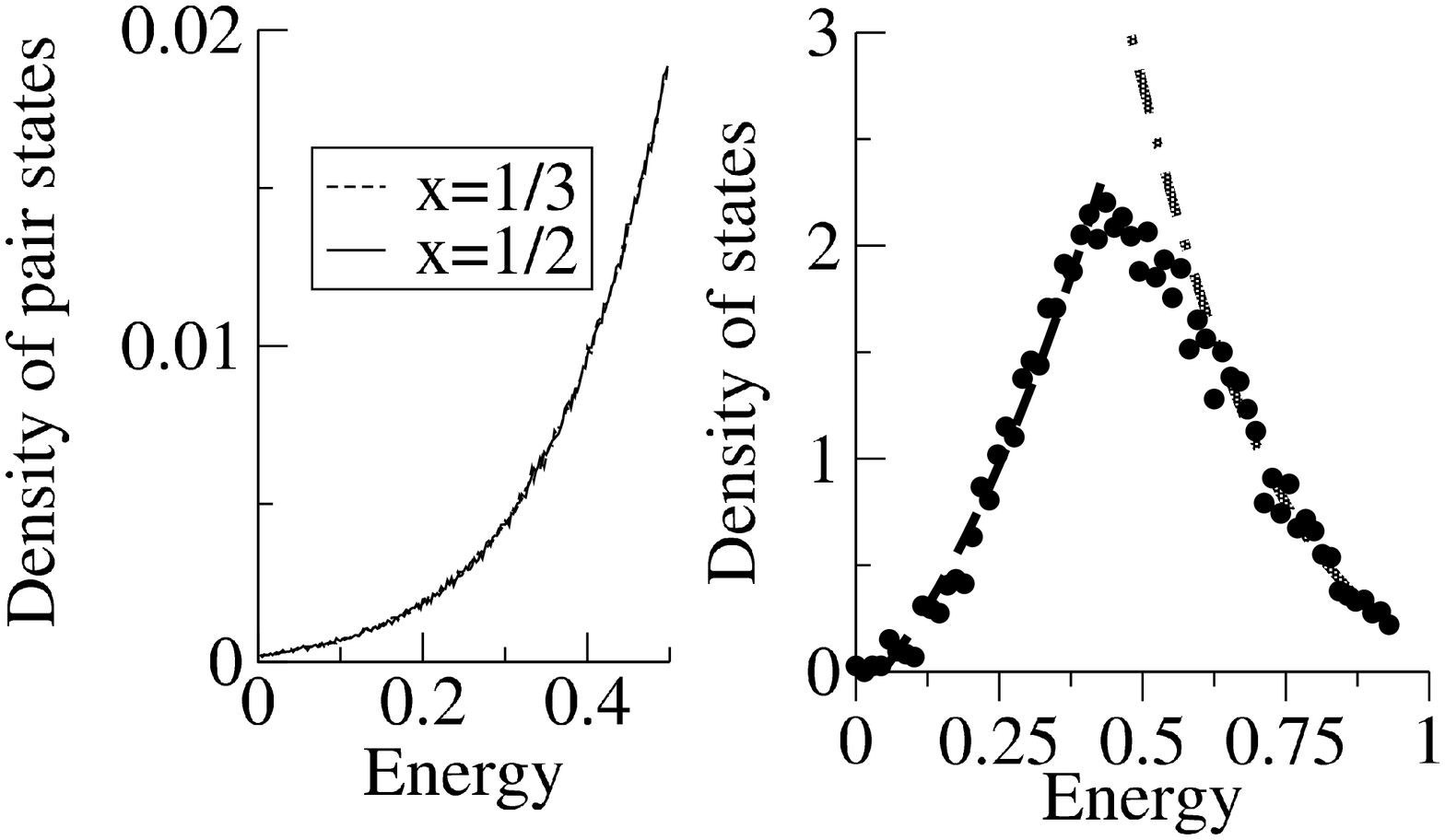}} \par}

\caption{\label{fig4} (a) Density of pair excitations at two different concentrations
of dopants.}

\label{fig5}(b) Distribution of local minima of full electronic energy (at
\( x=1/3 \)). The dashed line marks a parabolic dependence and the dash-dotted
curve is Gaussian.
\end{figure}

Another qualitative difference from the mean-field behavior was found in that
the density of pair excitations \( \nu _{p}(\epsilon ) \) does not remain constant
but tends to zero with \( \epsilon \rightarrow 0 \) (Fig. \ref{fig4}). Notably,
this result is concentration independent. It is of interest for the future studies,
to check also the low-energy behavior for the spectra of higher order excitations.

At least, the ``spectrum'' of local energy minima for the doping level \( x=1/3 \)
is shown in Fig. \ref{fig5}. Close to the value of absolute energy minimum
\( E_{min}^{abs} \) (which is proportional to the sample volume), this distribution
exhibits quadratic energy dependence (dashed line) and, farther from \( E_{min}^{abs} \),
it falls down by the Gaussian law (dotted line): \( \propto \exp \left[ -\alpha (E-E_{min}^{abs})^{2}\right]  \),
(while the overall distribution width is independent of the sample size and
relatively small).

\begin{figure}
{\centering \resizebox*{10cm}{!}{\includegraphics{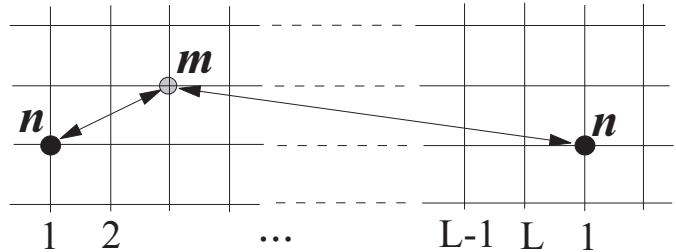}} \par}

\caption{\label{fig6} A lattice sample ``topologically closed'' in one dimension
(labeled from 1 to \protect\( L\protect \)), here two ``distances'' are defined
for each pair of points \protect\( \bf n\protect \) and \protect\( \bf m\protect \).}
\end{figure}

Besides the simplest cubic shape of the samples, we also examined the slabs
with \( L_{1}=L_{2}=6 \) and \( L_{3}=30 \). The essential features of single-particle
spectrum for them (overall extension and gap asymmetry) at \( x=1/3 \) were
found practically identical to those for cubic samples. At least, a modification
of the above slab configuration was considered, realizing a ``topologically
closed'' Euclidean bar (Fig. \ref{fig6}) where each vector \( (n_{1},n_{2},L+n_{3}) \)
is identified with \( (n_{1},n_{2},n_{3}) \). This form is appropriate for
direct simulations of current flow in a closed circuit under external electric
field applied along \( \bf e \)\( _{3} \). However in this case a special
care should be taken to assure the continuity of total energy at transitions
of particles through the ``topological'' interface: \( L\leftrightarrow 1 \),
which must be equivalent to any ``normal'' interface \( n_{3}\leftrightarrow n_{3}+1 \).
To this end, the interaction potential \( u(|{\bf n}-{\bf m}|) \) in Eq. (\ref{eq1})
should be replaced by the modified potential:
\begin{equation}
\label{eq5}
\tilde{u}({\bf n},{\bf m})=u(|{\bf n}-{\bf m}|)+u(|{\bf n}+L{\bf e}_{3}-{\bf m}|),
\end{equation}
(\( n_{3}<m_{3} \)), related to the two ``distances'' shown in Fig. \ref{fig6}.
Evidently, this modified interaction, Eq. (\ref{eq5}), which connects each
pair of sites in two possible ways, turns to the common Coulomb law in the limit
\( L\rightarrow \infty  \). Numeric simulations for a ``closed'' (6,6,30) slab with the modified interaction,
Eq. (\ref{eq5}), showed no significant difference in the single-particle spectrum,
compared to the similar ``open'' slab and usual interaction \( u(r) \). Thus,
the characteristics of the ground state and excitation spectra found from our
simulations are not sensitive to the sample size, shape and topology and should
correspond to the true thermodynamic limit of the strongly interacting disordered
system, Eq. (\ref{eq1}).

\section{Finite temperatures}

At finite temperatures, the system dynamics is determined by the thermally activated
hops between nearest neighbor sites. As usually, these hops are considered to
be controlled by the electron-phonon interaction in approximation of deformation
potential \cite{17}. If we consider only longitudinal phonons with Debye dispersion,
the transition rate between the neighboring sites \( \bf n \) and \( \bf n' \)
with the pair energy difference \( \epsilon ({\bf n},{\bf n'}=\epsilon ) \),
including the probabilities for both processes with phonon absorption (\( \epsilon >0 \))
and emission (\( \epsilon <0 \)), is given by a simple expression:
\begin{equation}
\label{eq6}
\gamma (\epsilon )=\gamma _{0}\frac{\epsilon -\epsilon _{D}\sin (\epsilon /\epsilon _{D})}{\exp (\epsilon /T)-1}\theta (\epsilon ^{2}_{D}-\epsilon ^{2})
\end{equation}
 where \( \epsilon _{D}=\hbar s/a \) is the Debye energy (for sound velocity
\( s \)) and \( \gamma _{0} \) is some constant proportional to the small
tunneling matrix element. The presence of the sine term in Eq. (\ref{eq6})
is due to the extremely short range of hops (by one lattice constant) and this
factor essentially reduces the transition amplitude at low energies, compared
to the usual case of electron-phonon interaction in metals.

If the mean interval of time between two consecutive transitions, in some volume
where the correlations are sensible, is much longer than the transition time
\( \tau _{tr} \) itself (this is reasonable for sufficiently small \( \gamma _{0} \)
and atomically fast \( \tau _{tr} \)), we can consider that only one transition
occurs at a time. Also, accordingly to the analysis by Knotek and Pollak \cite{18},
for so strictly localized states we can neglect transitions where more than
one electron participate. However, the difficulty with applying Eq. (\ref{eq6})
directly to our system consists in the fact that the transition energy for a
given pair of sites (in a given transition channel) is not \emph{a priori} defined
but depends on the overall configuration and is varied at transitions between
other sites. Since those transitions occur at random moments of time, the transition
rate for any given channel is itself a random function of time, correlated with
all other channels. In mathematical language, this means a realization of non-Markovian
branching random process \cite{19}. To manage this problem numerically, we
used the following algorithm.

For any initial configuration \( d(\{{\bf n}\}) \), \( c(\{{\bf n}\}) \),
Eq. (\ref{eq6}) defines the transition rates \( \gamma _{{\bf n},{\bf n'}}=\gamma [\epsilon ({\bf n},{\bf n'})] \)
for all appropriate pairs \( {\bf n}\rightarrow {\bf n'} \), and these alternatives
(the transition channels) can be considered independent. In this approach, only
one of the channels can be chosen for each consecutive transition, and this
choice is simulated by doing the independent statistical trials for random transition
times \( \tau _{{\bf n},{\bf n'}} \), accordingly to the rates \( \gamma _{{\bf n},{\bf n'}} \),
and choosing the shortest time from the trial outputs.

In each particular trial, a random number \( \xi  \), \( 0<\xi <1 \), is generated,
producing the associated random ``phase'' \( \phi =-\ln \xi  \). It relates
to the random output transition time \( \tau  \) at a constant transition rate
\( \gamma  \) through: \( \tau =\phi /\gamma  \). If \( \xi  \) is distributed
uniformly: \( P_{\xi }=\theta (\xi )\theta (1-\xi ) \), the related distribution
for the phase is: \( P_{\phi }=\exp (-\phi ) \), and hence the distribution
for transition times: \( P_{\tau }=\gamma \exp (-\gamma \tau ) \). The relation
between the random phase and transition time can be generalized to the case
when the transition rate \( \gamma  \) is not constant but \emph{a posteriori}
definite function of current time \( \gamma =\gamma (t) \):\begin{figure}
{\centering \resizebox*{12cm}{!}{\includegraphics{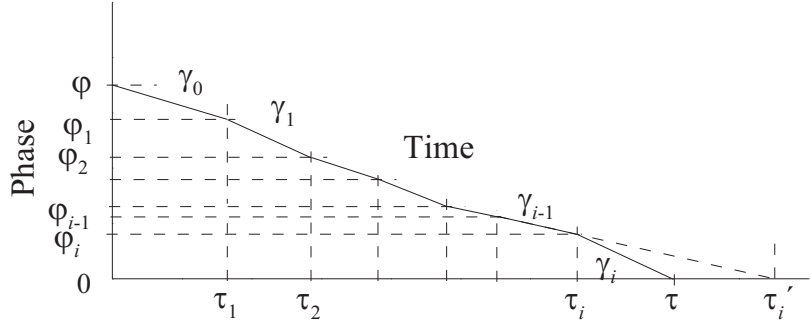}} \par}

\caption{\label{fig7}Realization of a transition through a certain channel with a random
apriori value of phase \protect\( \phi \protect \) and transition rates \protect\( \gamma _{0},...,\gamma _{i}\protect \)
which depend on time in a random way. The moment \protect\( \tau _{i}'\protect \)
corresponds to the \emph{virtual} value for this channel obtained at \protect\( \tau _{i-1}\protect \);
in fact this value was not realized because it ``lost'' the competition to
a shorter time \protect\( \tau _{i}\protect \) obtained at \protect\( \tau _{i-1}\protect \)
for other channel. The true transition time for this channel, \protect\( \tau \protect \),
is defined by the ``win'' of the output \protect\( \phi _{i}\gamma _{i}\protect \)
in the competition with outputs for all other channels at \protect\( \tau _{i}\protect \).}
\end{figure} 
\begin{equation}
\label{eq7}
\phi =\int ^{\tau }_{0}\gamma (t)dt
\end{equation}
 (in our specific case this function is stepwise, see Fig. \ref{fig7}). Since
the integrand in the r.h.s. of Eq. (\ref{eq7}) is nonnegative, there always
exists a single finite solution for the transition moment (Fig. \ref{fig7}).
This value is determined not only by the trial phase value \( \phi  \) but
also by all the intermediate times \( \tau _{i} \) and transition rates \( \gamma _{i}=\gamma (\tau _{i}<\tau <\tau _{i+1}) \)
expressing complicate correlations between the given transition and the preceding
ones. Each time interval \( \Delta \tau _{i}=\tau _{i+1}-\tau _{i} \) is determined
by the result of competition between the virtual values for all the channels
\( j \) possible after the moment \( \tau _{i} \):
\begin{equation}
\label{eq8}
\Delta \tau _{i}=\min _{j}(\phi _{i}^{(j)}/\gamma ^{(j)}_{i})
\end{equation}
 where \( \gamma _{i}^{(j)} \) is the transition rate in \( j \)-th channel
between the moments \( \tau _{i} \) and \( \tau _{i+1} \), and:
\begin{equation}
\label{eq9}
\phi ^{(i)}=\phi ^{(j)}-\int ^{\tau _{i}}_{\tau ^{(j)}}\gamma ^{(j)}(t)dt
\end{equation}
 is the ``residual phase at \( \tau _{i} \) for \( j \)-th channel. This
channel is opened at a certain time moment \( \tau ^{(j)} \), when it is given
the initial random phase value \( \phi ^{(j)} \), according to \( P_{\phi } \).
Then \( \phi ^{(j)} \) is consecutively reduced to \( \phi ^{(j)}_{i} \),
by Eq. (\ref{eq9}), at each intermediate transition, until such a moment \( \tau _{i} \)
is reached that the virtual value \( \phi ^{(j)}_{i}/\gamma ^{(j)}_{i} \) for
this channel ``wins'' the competition, Eq. (\ref{eq8}). Then the residual
phase value attained at \( \tau _{i+1}=\tau _{i}+\phi ^{(j)}_{i}/\gamma ^{(j)}_{i} \)
is just zero, corresponding to an exact solution of Eq. (\ref{eq7}) at \( \tau =\tau _{i+1} \).
After a particle has performed a transition through \( j \)-th channel, this
channel (together with the whole set of channels \( \bar{j} \), having common
initial site with \( j \)) is considered closed, and a number of new channels
is opened for all empty sites, neighbors to the new occupied site. Also, the
change of the system configuration produced by the transition in \( j \)-th
channel implies all the rates \( \gamma ^{(j')}_{i} \), \( j'\neq \bar{j} \),
to be changed for some new values \( \gamma _{i+1}^{(j')} \), and, after opening
of new channels with corresponding initial phases and transition rates, the
whole process is continued. At each \( i \)-th transition in the system, the
values of the energy transfer \( \epsilon _{i} \) (either positive or negative)
and of the time interval \( \Delta \tau _{i} \), past the preceding transition
are recorded.

It is important that, unlike a system of independent carriers, the whole passage
from opening to closure of a channel is not typically a pair process and it
is related to the overall density of many-body states.

\begin{figure}
{\centering \resizebox*{14cm}{!}{\includegraphics{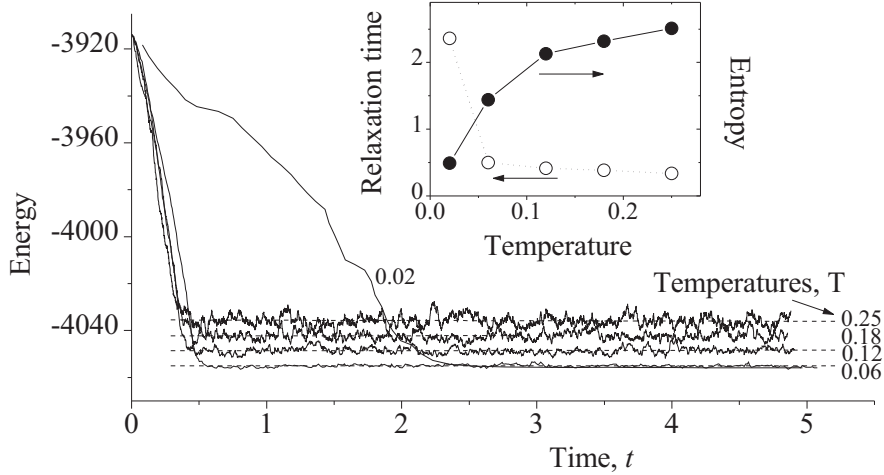}} \par}

\caption{\label{fig8} Temporal evolution of full energy in a cubic sample with \protect\( L=8\protect \)
and \protect\( x=1/3\protect \) at different temperatures. Time units are \protect\( \epsilon _{D}\gamma ^{-1}_{0}\times 10^{5}\protect \)
and energy units are \protect\( e^{2}/\kappa a\protect \). Inset: relaxation
time \protect\( t_{r}\protect \) and statistical entropy \protect\( S\protect \),
Eq. 11, vs temperature; a rapid increase of \protect\( t_{r}\protect \) below
\protect\( T\sim 0.05\protect \) indicates the freezing of a glassy system.}
\end{figure}

Examples of time records for the full energy as functions of full time \( \tau _{i}=\sum _{i'=1}^{i}\Delta \tau _{i'} \),
at different values of temperature \( T \) and concentration \( x=1/3 \) for
a cubic sample with \( L=8 \), are shown in Fig. \ref{fig8}. They all demonstrate
an initial rapid descent to the regime of dynamical equilibrium, within a certain
relaxation time \( t_{r} \). This time is almost independent of temperature
except at very low temperatures (below \( \sim 0.05 \), in our energy units
\( e^{2}/\kappa a \)) when \( t_{r} \) grows very rapidly (see inset to Fig.
\ref{fig8}). The latter can serve as an indication of the ``freezing'' process in the
glassy system, though, of course, there is no sharply defined critical temperature
in a finite size sample. In the equilibrium regime, both the mean energy \( \langle E\rangle  \)
and its dispersion \( \delta E=(\langle E^{2}\rangle -\langle E\rangle ^{2})^{1/2} \)
are obtained as certain functions of temperature.

Since the volume of our system is kept fixed and the average \emph{in time}
of total energy \( \langle E\rangle  \) as a function of temperature is known,
the specific heat \( C_{v} \) can be readily obtained by differentiation: \( C_{v}=\partial \langle E\rangle /\partial T \)
\cite{20}.

\begin{figure}
{\centering \resizebox*{10cm}{!}{\includegraphics{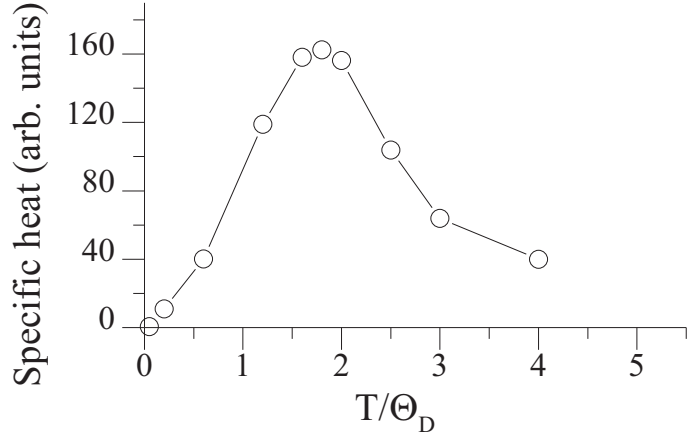}} \par}

\caption{\label{fig9} Specific heat vs temperature for the system corresponding to
Fig. 8. Points stand for the numerically calculated derivative \protect\( \partial E/\partial T\protect \),
and the solid line is a guide for the eye.}
\end{figure}

The corresponding numeric result is shown in Fig. \ref{fig9}. It displays a
linear behavior at low temperatures: \( C_{v}\propto T \), characteristic both
of the Fermi-liquid systems \cite{20} and of the glassy systems \cite{21}.

Another distinctive feature of the considered specific heat is a pronounced
maximum at \( T\sim 2\epsilon _{D} \), corresponding to saturation of the relaxation
channels above the Debye temperature. Through the relation:
\begin{equation}
\label{eq10}
S(T)=\int ^{T}_{0}\tau ^{-1}C_{v}(\tau )d\tau 
\end{equation}
 the thermodynamical entropy \( S(T) \) is fully determined by the function
\( \langle E(T)\rangle  \). On the other hand, \( S \) is also related to
\( \delta E \) \cite{20}:
\begin{equation}
\label{eq11}
S=ln[\nu _{m-b}(\langle E\rangle )\delta E],
\end{equation}
 which permits to estimate the total density of many-body states \( \nu _{m-b}(E) \),
very difficult in other approaches. From the comparison of Eqs. (\ref{eq10})
and (\ref{eq11}) at \( T\rightarrow 0 \), we conclude that \( \nu _{m-b}(E) \)
attains \emph{a finite value} near \( E_{min}^{abs} \), though, for a quantitative
accuracy, one perhaps needs more precise and detailed data on \( \delta E(T) \)
and \( \langle E(T)\rangle  \) than those in Figs. \ref{fig8}, \ref{fig9}.

Next, the temperature behavior of the hopping conductivity \( \sigma (T) \),
accordingly to the Einstein relation: \( \sigma =ne^{2}D/(k_{B}T) \) \cite{22},
can be estimated from that of the diffusion coefficient \( D \). Since all
the hops have a standard length \( a \), the diffusion coefficient \( D=a^{2}/(3\tau _{0}) \)
is fully determined by the mean lifetime \( \tau _{0} \) of a localized state,
and the latter is merely the inverse of the average number of hops per one particle
per unit time: \( \tau _{0}=N\lim _{i\rightarrow \infty }(\tau _{i}/i) \).
Then, from the plots of full number of hops \( i \) vs full time \( \tau _{i} \)
at different temperatures and the same concentration \( x=1/3 \) (they all
turn perfectly linear after the same relaxation time \( t_{r} \) as that for
energy, Fig. 9a), we deduced the values of \( D \), proportional to the slopes
\( di/d\tau _{i} \). Finally, the double logarithmic plot: \( \ln (\ln \sigma _{0}-\ln \sigma ) \)
vs \( \ln (1/T) \), Fig. 9b) (where the fitting parameter \( \sigma _{0} \)
was adjusted \begin{figure}
{\centering \resizebox*{14cm}{!}{\includegraphics{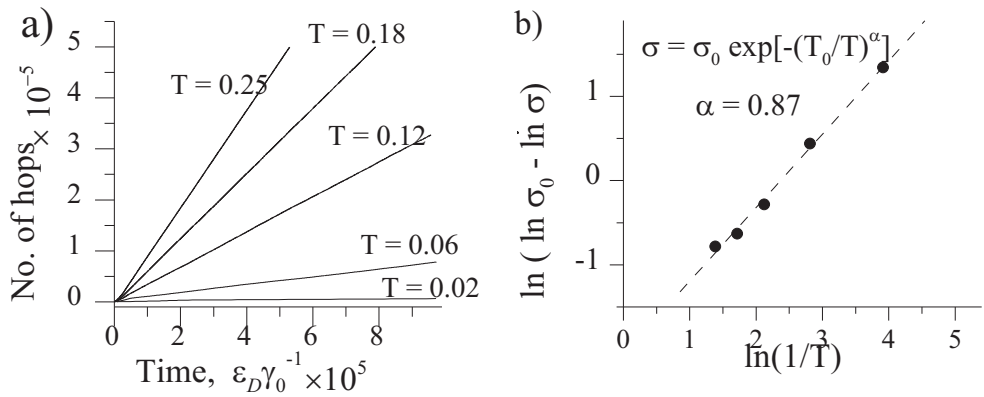}} \par}

\caption{\label{fig10}a) The slopes of linear functions \"{}full number of hops vs
full time\"{} provide the temperature dependence of the diffusion coefficient.
b) Double logarithmic plot for conductivity \protect\( \sigma \protect \) vs
\protect\( \ln \left( 1/T\right) \protect \) at \protect\( x=1/3\protect \).
The slope is different from the mean-field value \protect\( 1/2\protect \).}
\end{figure} to get the best linearity), permits to infer the modified hopping conductivity
law:
\begin{equation}
\label{eq12}
\sigma (T)=\sigma _{0}\exp [-(T_{0}/T)^{\alpha }],
\end{equation}
 with \( \alpha \approx 0.87 \) and reasonably low \( T_{0}\approx 0.1 \).
It is of interest to compare these figures with the mean-field values \cite{6}:
\( \alpha =1/2 \) and \( T_{0}\approx 2.1 \) (the latter is obtained using
the estimate for localization radius \( r_{0}\approx 0.4a \) by the Mott criterion
at \( x=1/3 \), see Introduction).

\section{Conclusions}

A model of strongly disordered lattice system with long-range Coulomb interactions
between localized charge carriers has been considered. The total electronic
energy is characterized by the presence of multiple metastable minima (including
the true ground state), and different types of excitation spectra over these
minima. A numeric procedure, accounting for all many-body correlations in finite
size samples, confirms the existence of Coulomb gap in the single-particle spectrum
and also provides corrections to the known mean-field theory results, as asymmetry
of the gap at non-half-filling, vanishing density of low energy pair excitations,
modified temperature exponent for hopping conductivity. The further analysis
of this model can involve direct simulations of current flow in a ``topologically
closed'' sample (Sec. 2) and formation of cluster states at finite tunneling
amplitudes between nearest neighbour sites with sufficiently small pair energies.

This research was supported by the Portuguese program PRAXIS XXI through the
project 2/2.1/FIS/302/94 and under personal Grants BPD 14226/97 (Yu.G.P.) and
BM/12717/97 (J. M. V. L.).

\end{document}